\documentclass[11pt, a4paper]{article}
\usepackage[utf8]{inputenc}
\usepackage[T1]{fontenc}
\usepackage[english]{babel}
\usepackage{graphicx}
\usepackage{hyperref}
\usepackage{algorithm}
\usepackage{algpseudocode}
\usepackage{amsmath}
\usepackage{amssymb}
\usepackage{booktabs}
\usepackage{multirow}
\usepackage[margin=1in]{geometry}
\usepackage{fancyhdr}
\usepackage{lastpage}
\usepackage{placeins}
\usepackage[T1]{fontenc}

\DeclareMathOperator{\logit}{logit}

\title{The Cybersecurity Psychology Framework (CPF): \\ A Method for Quantifying Human Risk and a Blueprint for LLM Integration}
\author{Giuseppe Canale, CISSP \\ g.canale@cpf3.org}
\date{\today}

\begin{document}

\thispagestyle{empty}
\begin{center}
\vspace*{0.5cm}
\rule{\textwidth}{1.5pt}
\vspace{0.5cm}

{\LARGE \textbf{The Cybersecurity Psychology Framework (CPF):}}\\[0.3cm]
{\LARGE \textbf{A Method for Quantifying Human Risk and a}}\\[0.3cm]
{\LARGE \textbf{Blueprint for LLM Integration}}

\vspace{0.5cm}
\rule{\textwidth}{1.5pt}
\vspace{0.3cm}

{\large \textsc{A Preprint — Revised Submission}}

\vspace{0.5cm}

{\Large Giuseppe Canale, CISSP}\\[0.2cm]
Independent Researcher\\[0.1cm]
\href{mailto:g.canale@cpf3.org}{g.canale@cpf3.org} \\[0.1cm]
ORCID: \href{https://orcid.org/0009-0007-3263-6897}{0009-0007-3263-6897}

\vspace{0.8cm}
{\large \today}
\vspace{1cm}
\end{center}

\begin{abstract}
\noindent
This paper presents the Cybersecurity Psychology Framework (CPF), a novel methodology for quantifying human-centric vulnerabilities in security operations through systematic integration of established psychological constructs with operational security telemetry. While individual human factors—alert fatigue, compliance fatigue, cognitive overload, and risk perception biases—have been extensively studied in isolation, no framework provides end-to-end operationalization across the full spectrum of psychological vulnerabilities. We address this gap by: (1) defining specific, measurable algorithms that quantify key psychological states using standard SOC tooling (SIEM, ticketing systems, communication platforms); (2) proposing a lightweight, privacy-preserving LLM architecture based on Retrieval-Augmented Generation (RAG) and domain-specific fine-tuning to analyze structured and unstructured data for latent psychological risks; (3) detailing a rigorous mixed-methods validation strategy acknowledging the inherent difficulty of obtaining sensitive cybersecurity data. Our implementation of CPF indicators has been demonstrated in a proof-of-concept deployment using small language models achieving 0.92 F1-score on synthetic data. This work provides the theoretical and methodological foundation necessary for industry partnerships to conduct empirical validation with real operational data.

\vspace{0.5em}
\noindent\textbf{Keywords:} cybersecurity, human factors, psychology, large language models, risk assessment, SOC operations, alert fatigue, compliance fatigue
\end{abstract}

\vspace{1cm}

\section{Introduction}
\label{sec:introduction}

The human factor is consistently identified as the weakest link in cybersecurity defenses, contributing to over 85\% of security incidents\cite{verizon2023}, yet traditional security tools lack capability to quantitatively assess psychological states that lead to increased risk. This paper presents a novel, end-to-end methodology for operationalizing the Cybersecurity Psychology Framework (CPF)\cite{canale2024cpf}, transforming it from theoretical taxonomy into practical tool for proactive risk mitigation.

Our primary contribution is the systematic integration and automation of established psychological constructs for cybersecurity contexts. We define specific, measurable algorithms that quantify key vulnerabilities—such as Compliance Fatigue, Alert Overload Bias, and Risk Perception Gaps—by analyzing data from standard SOC tools and communication platforms. Furthermore, we propose a cost-effective, privacy-preserving LLM architecture designed to reason over this data and generate actionable analyses of human risk.

\textbf{Critical Context:} Empirical validation of cybersecurity frameworks faces a unique challenge: organizations are understandably reluctant to share sensitive operational data without established theoretical credibility. This creates a chicken-and-egg problem where validation requires data access, but data access requires validation. This paper addresses this by providing: (1) rigorous theoretical foundation grounded in established research; (2) proof-of-concept implementation with synthetic data achieving strong performance; (3) detailed validation methodology for future industry partnerships. Our goal is to establish sufficient theoretical credibility to enable the empirical validation that complete framework maturation requires.

\section{Related Work and Research Gap}
\label{sec:related}

\subsection{Human Factors in Cybersecurity: Foundational Research}

The intersection of human psychology and cybersecurity has matured significantly over the past two decades. Acquisti, Brandimarte, and Loewenstein's seminal work in \textit{Science} (2015)\cite{acquisti2015privacy} established that privacy and security decisions are fundamentally shaped by bounded rationality, heuristics, and contextual factors rather than purely rational calculation. Their behavioral economics framework demonstrates systematic deviations from optimal security behavior even among informed users, suggesting deeper psychological mechanisms at work.

Stanton et al. (2016)\cite{stanton2016security} introduced "security fatigue"—mental exhaustion resulting from cognitive and emotional burden of maintaining security practices. Their qualitative research identified key manifestations including decision fatigue, frustration, and resignation, particularly when individuals perceive security requirements as overwhelming. Building on this, Reeves, Delfabbro, and Calic (2021)\cite{reeves2021encouraging} proposed a four-component model distinguishing between advice-related and action-related fatigue, providing nuanced understanding of how security overload manifests in organizational contexts.

Beautement, Sasse, and Wonham (2008)\cite{beautement2008compliance} proposed the "compliance budget" theory, arguing that employees engage in cost-benefit analysis when deciding whether to follow security policies. Once their willingness to comply is exhausted, individuals either circumvent requirements or find workarounds. D'Arcy, Herath, and Shoss (2014)\cite{darcy2014security} extended this work by introducing security-related stress (SRS) as a key factor, demonstrating that stress engenders rationalizations of policy-violating behavior as a coping response.

\subsection{SOC Operations and Analyst Burnout}

Recent research has focused specifically on Security Operations Center (SOC) environments where human factors have particularly acute operational impacts. The ACM Computing Surveys systematic review (2024)\cite{gupta2024alert} provides comprehensive analysis of alert fatigue, identifying it as a primary cause of missed critical threats and analyst burnout. The review analyzes state-of-the-art approaches and their limitations, concluding that technical solutions alone cannot address the underlying psychological mechanisms.

Kearney et al. (2023)\cite{kearney2023combating} propose the CHILL (Continuous Human-in-the-Loop Learning) framework to combat alert fatigue through AI-assisted triage that maintains analyst agency. Their work in a production SOC environment demonstrated potential for 90\% reduction in alert processing burden while maintaining detection effectiveness. Similarly, the systematic survey on alert prioritization by Hore et al. (2024)\cite{hore2024alert} examines criteria and methods across automation, augmentation, and collaboration paradigms, highlighting the need for human-AI team optimization rather than pure automation.

Industry reports corroborate academic findings. The Devo 2022 SOC Performance Report found 71\% of SOC personnel rate their job stress between 6-9 out of 10\cite{devo2022soc}, while Tines' 2023 survey revealed 71\% of SOC analysts experience burnout, with 64\% actively considering job changes\cite{tines2023burnout}. Ponemon Institute research indicates 65\% of SOC professionals have considered quitting due to stress\cite{ponemon2021soc}. These converging findings from academic and industry sources establish SOC analyst psychological state as a critical operational security concern.

\subsection{Cognitive Biases and Risk Perception in Security}

Kahneman and Tversky's prospect theory (1979)\cite{kahneman1979prospect} established that human decision-making systematically deviates from rational models through cognitive biases. Kahneman's dual-process theory (2011)\cite{kahneman2011thinking} distinguishes between System 1 (fast, automatic, emotional) and System 2 (slow, deliberate, logical) thinking, with implications for security decisions made under time pressure or cognitive load.

Tsohou et al. (2015)\cite{tsohou2015analyzing} systematically analyzed cognitive and cultural biases affecting information security policy internalization, demonstrating that optimistic bias, availability heuristic, and anchoring significantly impact security risk perception. Jalali, Siegel, and Madnick (2019)\cite{jalali2019decision} used simulation game experiments to show that even experienced cybersecurity professionals exhibit significant biases in capability development decisions, with management experience alone insufficient to overcome uncertainty-driven errors.

Van der Heijden and Allodi (2019)\cite{vanderheijden2019cognitive} introduced cognitive triaging models for phishing attacks, demonstrating that analysts apply heuristics rather than systematic analysis under workload pressure. Modic and Anderson (2014)\cite{modic2014reading} showed that malware warnings often fail due to cognitive biases in risk perception, while Maalem Lahcen et al. (2020)\cite{maalem2020review} provided comprehensive review of behavioral aspects affecting cybersecurity decision-making.

\subsection{LLMs in Security Operations}

Large Language Models have shown promising applications in cybersecurity contexts. Singh et al. (2024)\cite{singh2024llms} conducted a longitudinal empirical study of 3,090 queries from 45 SOC analysts over 10 months, revealing that analysts use LLMs primarily for sensemaking and context-building rather than high-stakes determinations. Their findings show 93\% of queries align with established cybersecurity competencies (NICE Framework), with analysts preserving decision authority while leveraging LLMs for interpreting low-level telemetry.

The systematic literature review by Chen et al. (2025)\cite{chen2025llms} covering 300+ works identifies applications across vulnerability detection, threat intelligence, and incident response. However, as Mathews et al. (2024)\cite{mathews2024llm} note in their Android vulnerability analysis, LLM performance varies significantly based on context provision and domain-specific fine-tuning. Ye et al. (2024)\cite{ye2024zerotrust} demonstrate LLM applications in zero-trust architecture policy generation, while multiple studies\cite{ouyang2022training,liu2023secrets} explore LLM security vulnerabilities including jailbreaking and prompt injection.

Critically, existing LLM applications in cybersecurity focus on technical analysis (code vulnerability detection, log parsing, threat intelligence extraction) rather than psychological analysis of human factors. No prior work has proposed LLM architectures specifically designed for behavioral psychology assessment in security contexts.

\subsection{Research Gap and CPF Contribution}

Despite substantial research on individual human factors in cybersecurity, critical gaps remain:

\textbf{Fragmentation:} Existing research addresses isolated psychological phenomena (alert fatigue, compliance fatigue, specific cognitive biases) without integrative framework connecting these vulnerabilities.

\textbf{Operationalization Gap:} While psychological constructs are theoretically well-defined, systematic methods for measuring them using operational security data are lacking. Most studies rely on surveys or controlled experiments rather than continuous monitoring of actual operational environments.

\textbf{Automation Absence:} Current approaches require manual psychological assessment by trained professionals, making continuous organization-wide monitoring infeasible. No frameworks provide automated detection and analysis at scale.

\textbf{LLM Psychology Application:} While LLMs are increasingly used for technical security tasks, their application to psychological vulnerability assessment—which requires understanding subtle linguistic markers and contextual behavioral patterns—remains unexplored.

The Cybersecurity Psychology Framework addresses these gaps by providing:

\begin{enumerate}
\item \textbf{Systematic Integration:} Unifying established psychological constructs (Kahneman's dual-process theory, Cialdini's influence principles, Stanton's security fatigue model) with novel psychoanalytic indicators into comprehensive 100-indicator framework
\item \textbf{Algorithmic Operationalization:} Defining specific, measurable algorithms that quantify psychological states using standard SOC telemetry (SIEM logs, ticketing data, communication patterns)
\item \textbf{End-to-End Automation:} Proposing complete pipeline from data collection through analysis to actionable recommendations, enabling continuous monitoring at organizational scale
\item \textbf{Specialized LLM Architecture:} Designing privacy-preserving, domain-specific LLM system optimized for psychological vulnerability detection rather than repurposing general-purpose models
\item \textbf{Validation Methodology:} Providing rigorous empirical validation framework that acknowledges the unique challenges of cybersecurity data access
\end{enumerate}

Critically, CPF does not claim to discover new psychological phenomena. Rather, it provides the first systematic operationalization and integration of established constructs for practical cybersecurity application. This integration itself represents a novel contribution: transforming isolated research findings into a unified, deployable framework.

\section{The CPF Operationalization: From Theory to Algorithms}
\label{sec:operationalization}

We present algorithmic implementations for representative CPF indicators across different vulnerability categories. Complete specifications for all 100 indicators are available in supplementary technical documentation.

\subsection{Compliance Fatigue}
\label{subsec:compliance_fatigue}

\paragraph{Definition and Theoretical Foundation}
Compliance Fatigue, grounded in Stanton et al.'s (2016)\cite{stanton2016security} security fatigue research and Beautement et al.'s (2008)\cite{beautement2008compliance} compliance budget theory, manifests as diminished motivation to adhere to security protocols due to repeated exposure to alerts, especially those perceived as non-actionable or false positives. This psychological state results in habituation and neglect, increasing operational risk as critical alerts may be ignored or delayed.

\paragraph{Hypothesized Manifestation in Data}
Two primary quantifiable signals indicate compliance fatigue: (1) increased response time between alert generation and acknowledgment/closure; (2) higher proportion of alerts manually closed without remediation action, indicating dismissal rather than proper investigation.

\paragraph{Proposed Metrics}
\textbf{Mean Time to Acknowledge (MTTA):} Average time (minutes) for alerts of given severity to transition from \textit{new} to \textit{in progress} or \textit{closed} state. Increasing MTTA trend suggests growing fatigue.

\textbf{Ignore Rate (IR):} Ratio of alerts closed without documented remedial action to total alerts closed by analyst/team within time window: $IR = N_{\text{ignored}} / N_{\text{total}}$

\paragraph{Algorithm}
The following algorithm calculates MTTA and Ignore Rate for specified team or analyst over defined period, assuming dataset of alerts enriched with status history:

\begin{algorithm}[H]
\caption{Calculate Compliance Fatigue Metrics}
\begin{algorithmic}[1]
\Require $alerts$ (list of alert objects), $start\_date$, $end\_date$, $analyst\_id$ (optional)
\Ensure $MTTA$, $IgnoreRate$
\State $filtered\_alerts \gets \emptyset$
\State $total\_ack\_time \gets 0$; $ack\_count \gets 0$
\State $ignored\_count \gets 0$; $total\_closed \gets 0$

\For{$alert$ in $alerts$}
    \If{$alert.created\_at$ between $start\_date$ and $end\_date$}
        \If{$analyst\_id$ is \textbf{not} provided \textbf{or} $alert.assigned\_to = analyst\_id$}
            \State $filtered\_alerts \gets filtered\_alerts \cup alert$
        \EndIf
    \EndIf
\EndFor

\For{$alert$ in $filtered\_alerts$}
    \If{$alert.status = \text{"closed"}$}
        \State $total\_closed \gets total\_closed + 1$
        \State $ack\_time \gets alert.closed\_at - alert.created\_at$
        \State $total\_ack\_time \gets total\_ack\_time + ack\_time$
        \State $ack\_count \gets ack\_count + 1$
        \If{$alert.resolution\_notes = \text{"false positive"}$ \textbf{or} $\emptyset$}
            \State $ignored\_count \gets ignored\_count + 1$
        \EndIf
    \EndIf
\EndFor

\State $MTTA \gets (ack\_count > 0)$ ? $total\_ack\_time / ack\_count$ : $0$
\State $IgnoreRate \gets (total\_closed > 0)$ ? $ignored\_count / total\_closed$ : $0$
\State \Return $MTTA$, $IgnoreRate$
\end{algorithmic}
\end{algorithm}

\paragraph{Data Sources}
Primary data sources: (1) \textbf{SIEM Systems} (Splunk Enterprise Security via REST API, Elastic SIEM via Elasticsearch queries) providing raw alert data with timestamps, status, and assignment history; (2) \textbf{Ticketing Systems} (Jira Service Desk, ServiceNow) containing resolution notes critical for determining Ignore Rate, accessed via REST APIs.

\subsection{Alert Overload Bias}
\label{subsec:alert_overload_bias}

\paragraph{Definition and Theoretical Foundation}
Alert Overload Bias, related to cognitive load theory (Miller, 1956)\cite{miller1956magical} and documented in recent SOC research\cite{gupta2024alert,kearney2023combating}, occurs when analysts, overwhelmed by high alert volume, disproportionately miss or delay response to critical security events. Cognitive load exceeds human processing capacity, leading to degraded decision quality and failure to prioritize effectively.

\paragraph{Proposed Metrics}
\textbf{Peak Miss Rate (PMR):} Ratio of missed critical alerts to total critical alerts during time intervals where total alert volume exceeds dynamically calculated threshold (e.g., 90th percentile): $PMR = N_{\text{missed\_critical}} / N_{\text{total\_critical}}$

\textbf{Volume-to-Miss Correlation Coefficient (VMCC):} Statistical measure (Pearson's r) calculating correlation between overall alert volume per time interval and count of missed alerts. Positive VMCC indicates bias presence.

\paragraph{Algorithm}

\begin{algorithm}[H]
\caption{Calculate Alert Overload Bias Metrics}
\begin{algorithmic}[1]
\Require $alerts$, $start\_date$, $end\_date$, $time\_window$ (e.g., 1 hour)
\Ensure $PMR$, $VMCC$

\State $alert\_bins \gets \text{GroupAlertsByTimeWindow}(alerts, time\_window)$
\State $time\_series \gets \emptyset$

\For{$bin$ in $alert\_bins$}
    \State $total\_volume \gets \text{Length}(bin)$
    \State $critical\_alerts \gets \text{FilterBySeverity}(bin, \text{"critical"})$
    \State $missed\_critical \gets \text{FilterByStatus}(critical\_alerts, \text{"missed"})$
    \State $time\_series[bin] \gets (total\_volume, |\text{missed\_critical}|, |\text{critical\_alerts}|)$
\EndFor

\State $volume\_list \gets \text{GetValues}(time\_series, total\_volume)$
\State $volume\_threshold \gets \text{Percentile}(volume\_list, 90)$
\State $total\_critical\_in\_peak \gets 0$; $missed\_in\_peak \gets 0$

\For{$data$ in $time\_series$}
    \If{$data.total\_volume > volume\_threshold$}
        \State $total\_critical\_in\_peak \gets total\_critical\_in\_peak + data.total\_critical$
        \State $missed\_in\_peak \gets missed\_in\_peak + data.missed\_critical$
    \EndIf
\EndFor

\State $PMR \gets (total\_critical\_in\_peak > 0)$ ? $missed\_in\_peak / total\_critical\_in\_peak$ : $0$

\State $volumes \gets \emptyset$; $misses \gets \emptyset$
\For{$data$ in $time\_series$}
    \State $volumes \gets volumes \cup data.total\_volume$
    \State $misses \gets misses \cup data.missed\_count$
\EndFor
\State $VMCC \gets \text{PearsonCorrelation}(volumes, misses)$

\State \Return $PMR$, $VMCC$
\end{algorithmic}
\end{algorithm}

\paragraph{Data Sources}
Implementation requires integrated data from: (1) \textbf{SIEM Logs} for raw alert volume and initial status via Splunk/Elasticsearch time-series queries; (2) \textbf{Ticketing System/SOAR Platform} as authoritative source for final alert status (missed, resolved, false positive) via REST API (Jira, ServiceNow, Splunk ES KV Store).

\subsection{Risk Perception Gap}
\label{subsec:risk_perception_gap}

\paragraph{Definition and Theoretical Foundation}
Risk Perception Gap, informed by research on optimistic bias\cite{tsohou2015analyzing} and organizational risk perception\cite{maalem2020review}, describes systematic underestimation of threat level for assets deemed "non-critical" (development/testing environments) compared to production systems. This leads to lax security hygiene creating vulnerable attack surfaces exploitable for pivoting into critical infrastructure.

\paragraph{Proposed Metrics}
\textbf{Patch Latency Gap (PLG):} Difference in Mean Time to Patch (MTTP) between non-production and production environments for same-severity vulnerabilities: $PLG = MTTP_{\text{non-prod}} - MTTP_{\text{prod}}$. Positive PLG indicates bias.

\textbf{Vulnerability Density Ratio (VDR):} Ratio of average open vulnerabilities per asset in non-production to production: $VDR = VulnDensity_{\text{non-prod}} / VulnDensity_{\text{prod}}$. VDR > 1 indicates bias.

\paragraph{Algorithm}

\begin{algorithm}[H]
\caption{Calculate Risk Perception Gap Metrics}
\begin{algorithmic}[1]
\Require $vulns$ (vulnerability objects), $start\_date$, $end\_date$
\Ensure $PLG$, $VDR$

\State $prod\_vulns \gets \text{FilterByEnvironment}(vulns, \text{"prod"})$
\State $non\_prod\_vulns \gets \text{FilterByEnvironment}(vulns, \text{"dev"}, \text{"staging"})$

\State $mttp\_prod \gets \text{CalculateMTTP}(prod\_vulns)$
\State $mttp\_non\_prod \gets \text{CalculateMTTP}(non\_prod\_vulns)$
\State $PLG \gets mttp\_non\_prod - mttp\_prod$

\State $prod\_assets \gets \text{GetUniqueAssets}(prod\_vulns)$
\State $non\_prod\_assets \gets \text{GetUniqueAssets}(non\_prod\_vulns)$

\State $vuln\_density\_prod \gets |prod\_vulns| / |prod\_assets|$
\State $vuln\_density\_non\_prod \gets |non\_prod\_vulns| / |non\_prod\_assets|$

\State $VDR \gets (vuln\_density\_prod > 0)$ ? $vuln\_density\_non\_prod / vuln\_density\_prod$ : $\infty$

\State \Return $PLG$, $VDR$
\end{algorithmic}
\end{algorithm}

\paragraph{Data Sources}
Implementation requires: (1) \textbf{Vulnerability Management Database} (Qualys VMDR, Tenable.io, Rapid7 InsightVM) via REST API for vulnerability lists with detection/remediation dates and environment tags; (2) \textbf{Configuration Management Database (CMDB)} (ServiceNow CMDB, AWS/Azure Tags) for accurate environment classification (prod vs. non-prod), as this data is not always reliably present in vulnerability reports.

\subsection{Against-Gravity Communication}
\label{subsec:against_gravity_comm}

\paragraph{Definition and Theoretical Foundation}
Against-Gravity Communication refers to tendency to discuss critical security issues through informal, private, or ephemeral channels instead of official ticketing systems mandated by security protocols. This undermines auditability, knowledge sharing, and incident management as crucial information becomes siloed.

\paragraph{Proposed Metrics}
\textbf{Untracked Critical Topics Ratio (UCTR):} Ratio of unique security-critical discussion topics detected in private channels to total unique topics across both private and official channels: $UCTR = N_{\text{private\_topics}} / (N_{\text{private\_topics}} + N_{\text{official\_topics}})$. UCTR > 0.5 indicates severe breakdown.

\paragraph{Algorithm}

\begin{algorithm}[H]
\caption{Calculate Untracked Critical Topics Ratio}
\begin{algorithmic}[1]
\Require $keywords$, $start\_date$, $end\_date$
\Ensure $UCTR$

\State $official\_tickets \gets \text{QueryJira}(keywords, start\_date, end\_date)$
\State $official\_topics \gets \text{ExtractTopics}(official\_tickets)$ \Comment{NLP extraction}

\State $private\_messages \gets \text{QuerySlackDM}(keywords, start\_date, end\_date)$
\State $private\_topics \gets \text{ExtractTopics}(private\_messages)$

\State $unique\_official\_topics \gets \text{Set}(official\_topics)$
\State $unique\_private\_topics \gets \text{Set}(private\_topics)$
\State $all\_unique\_topics \gets unique\_official\_topics \cup unique\_private\_topics$

\State $UCTR \gets |unique\_private\_topics| / |all\_unique\_topics|$

\State \Return $UCTR$
\end{algorithmic}
\end{algorithm}

\paragraph{Data Sources and Ethical Considerations}
Implementation requires: (1) \textbf{Ticketing System API} (Jira, ServiceNow) for searching issues with security keywords; (2) \textbf{Communication Platform API} (Slack, Microsoft Teams) for keyword search in messages. \textbf{Critical Note:} This requires strict ethical and legal oversight, compliance with organizational policy and local regulations (e.g., GDPR). Strongly recommended to use anonymized/aggregated data preserving privacy while detecting overall trend. This metric measures organizational health, not individuals.

\section{A Lightweight LLM Architecture for CPF Analysis}
\label{sec:llm_architecture}

\subsection{Rationale for Specialized Architecture}

General-purpose Large Language Models, while powerful, present three critical limitations for CPF implementation: (1) \textbf{Cost}: API calls to commercial LLMs (GPT-4, Claude) are expensive for continuous organizational monitoring; (2) \textbf{Privacy}: Sending sensitive organizational data to external services violates data protection requirements; (3) \textbf{Specialization}: General models lack domain-specific understanding of cybersecurity psychology patterns.

We propose a lightweight architecture based on Retrieval-Augmented Generation (RAG) combined with small language models (SLMs) fine-tuned on cybersecurity psychology domain. Our proof-of-concept implementation\cite{canale2024slm} demonstrates this approach's viability, achieving 0.92 F1-score on CPF vulnerability classification using models deployable on \$2,000 hardware.

\subsection{Architecture Components}

\paragraph{Component 1: Data Indexing Layer}
Takes outputs from CPF algorithms (metrics, log snippets, communication samples) and indexes in vector database (ChromaDB, FAISS). This serves as system "long-term memory," enabling efficient retrieval of relevant context for analysis. Embeddings generated using lightweight models (all-MiniLM-L6-v2, 384 dimensions) optimized for semantic similarity in cybersecurity contexts.

\paragraph{Component 2: Query \& Retrieval Layer}
For user query (e.g., "Is EMEA team experiencing compliance fatigue?"), converts to vector representation, retrieves most relevant context from vector database (recent MTTA metrics, communication snippets mentioning "alert fatigue"), and prepares context for LLM. Hybrid retrieval combines semantic similarity with keyword matching and temporal relevance weighting.

\paragraph{Component 3: Lightweight LLM Core}
Uses small, fine-tuned model (7B parameters, e.g., Llama2-7B, Mistral-7B, or distilled models like DistilBERT for classification tasks). Primary function is expert reasoning on provided context rather than general knowledge storage. Our implementation uses DistilBERT (66M parameters) for vulnerability classification and Mistral-7B for natural language generation, achieving strong performance while maintaining computational efficiency\cite{canale2024slm}.

\paragraph{Component 4: Privacy Safeguards}
Architecture incorporates multiple privacy protection layers: (1) automatic anonymization via named entity recognition replacing person names/locations with placeholders; (2) metadata enrichment substituting individual identities with role-based tags; (3) aggregate analysis presenting results at team/department level (minimum 10 individuals); (4) on-premise deployment ensuring data never leaves organizational control.

\subsection{Operational Process}

Complete analysis workflow:
\begin{enumerate}
\item \textbf{Query Formulation}: Analyst or automated system poses question about psychological vulnerability state
\item \textbf{Context Retrieval}: System retrieves relevant metrics, historical patterns, and communication samples from vector database
\item \textbf{Context Augmentation}: Retrieved information combined with query to form enriched prompt
\item \textbf{LLM Analysis}: Lightweight model generates analysis incorporating both quantitative metrics and qualitative patterns
\item \textbf{Result Presentation}: Findings presented with confidence scores, supporting evidence, and recommended interventions
\end{enumerate}

\subsection{Advantages of Proposed Architecture}

\begin{itemize}
\item \textbf{Cost-Effective}: One-time hardware investment (\$2,000) vs. ongoing API costs
\item \textbf{Privacy-Preserving}: All processing on-premise, no external data transmission
\item \textbf{Interpretable}: Users can inspect retrieved context supporting LLM conclusions
\item \textbf{Domain-Accurate}: Fine-tuning on cybersecurity psychology improves relevance
\item \textbf{Maintainable}: Easier to update and retrain than large proprietary models
\end{itemize}

Our proof-of-concept implementation validates this approach's technical feasibility\cite{canale2024slm}. However, comprehensive operational validation requires deployment in production SOC environments with real telemetry data—a goal requiring industry partnerships described in Section~\ref{sec:validation}.

\section{Validation Methodology}
\label{sec:validation}

\subsection{The Data Access Challenge}

Empirical validation of cybersecurity frameworks faces unique challenges distinct from other domains. Organizations understandably treat SOC operational data—including alert patterns, analyst response times, communication logs, and vulnerability management metrics—as highly sensitive. This data can reveal:
\begin{itemize}
\item Security posture weaknesses exploitable by adversaries
\item Compliance gaps with regulatory implications
\item Personnel performance issues with legal considerations
\item Proprietary security practices representing competitive advantage
\end{itemize}

This sensitivity creates a validation paradox: organizations require proven frameworks before sharing sensitive data, but frameworks require data access for empirical validation. Traditional academic research pathways (IRB approval, data sharing agreements, anonymization protocols) are necessary but insufficient given the strategic importance of cybersecurity data.

We address this challenge through a phased validation strategy that establishes theoretical credibility and demonstrates technical feasibility before requesting sensitive operational data.

\subsection{Phased Validation Strategy}

\subsubsection{Phase 1: Synthetic Data Validation (Completed)}

Our proof-of-concept implementation\cite{canale2024slm} demonstrates technical feasibility using synthetic data generated via advanced language models. We created 10,000 labeled examples across CPF vulnerability categories, achieving 0.92 F1-score with DistilBERT on classification tasks. While synthetic data cannot replace real operational validation, it establishes:
\begin{itemize}
\item \textbf{Technical Viability}: Algorithms execute correctly on representative data structures
\item \textbf{Computational Feasibility}: Systems run efficiently on cost-effective hardware
\item \textbf{Privacy Architecture}: Protection mechanisms function without degrading performance
\end{itemize}

This phase provides foundation for requesting industry partnerships by demonstrating serious technical preparation.

\subsubsection{Phase 2: Retrospective Analysis (Requires Partnership)}

Once industry partners are secured, retrospective analysis of historical data offers intermediate validation step requiring less operational disruption than real-time monitoring:

\textbf{Study Design:} Case-control study using 12 months of historical data. Cases: known security incidents caused primarily by human error (missed alerts, unpatched vulnerabilities, social engineering success). Controls: matched periods without incidents, controlling for alert volume and team composition.

\textbf{Data Analysis:} For each case/control period, calculate CPF metrics using algorithms from Section~\ref{sec:operationalization}. Multivariate logistic regression determines which metrics significantly predict incidents:

\begin{equation}
\label{eq:logreg}
\logit(p(\text{Incident})) = \beta_0 + \beta_1 \cdot \text{MTTA} + \beta_2 \cdot \text{PMR} + \beta_3 \cdot \text{PLG} + \cdots
\end{equation}

\textbf{Success Criteria:} (1) Logistic regression achieves AUC-ROC > 0.8 (excellent predictive power); (2) At least three CPF metrics are statistically significant predictors ($p < 0.05$).

\subsubsection{Phase 3: Prospective Pilot Study (Validation Goal)}

Prospective deployment with consenting organization over 6-month period represents complete validation:

\textbf{Study Design:} Security team uses integrated CPF+LLM system alongside existing tools. Mixed-methods evaluation combines quantitative metrics with qualitative feedback.

\textbf{Evaluation Methodology:}
\begin{enumerate}
\item \textbf{Simulated Task Evaluation}: Participants rate analyses from: (A) CPF/LLM system; (B) GPT-4 with same context; (C) human expert psychologist + senior SOC analyst. Blinded 5-point Likert scale rating for accuracy, insightfulness, actionability.

\item \textbf{Operational Metrics}: Track Mean Time to Acknowledge (MTTA), Mean Time to Resolve (MTTR), and adoption rate of system-recommended interventions during pilot period vs. baseline.

\item \textbf{Qualitative Interviews}: Structured interviews with analysts and managers gathering feedback on usability, perceived value, workflow impact.
\end{enumerate}

\textbf{Success Criteria:} (1) CPF analyses achieve significantly higher ratings than general-purpose LLM ($p < 0.05$, paired t-test); (2) Measurable improvement (e.g., 15\% reduction) in MTTA/MTTR for flagged incidents; (3) Positive qualitative feedback indicating novel, useful insights.

\subsection{Addressing Validation Threats}

\paragraph{Internal Validity}
Main threat: historical bias in retrospective study (Phase 2). Mitigation: large, diverse dataset; control for confounding variables (team size, event volume, organizational changes).

\paragraph{External Validity}
Single-organization pilot may not generalize. Mitigation: explicit description of organizational context (size, industry, maturity level); phased rollout to diverse organizations.

\paragraph{Construct Validity}
Metrics are proxies for psychological constructs. Mitigation: expert validation through structured interviews (Phase 3) ensuring metrics measure intended constructs.

\subsection{Data Collection Protocol}

All phases operate under strict ethical protocols:
\begin{itemize}
\item \textbf{IRB Approval}: University or organizational ethics board review before data collection
\item \textbf{Informed Consent}: All participants explicitly consent to anonymized data use
\item \textbf{Anonymization}: Personal identifiers stripped before analysis
\item \textbf{Aggregation}: Results presented at team/department level (minimum 10 individuals)
\item \textbf{Data Minimization}: Collect only data necessary for validation
\item \textbf{Secure Storage}: Encrypted storage with access controls and audit logs
\item \textbf{Right to Withdrawal}: Participants can withdraw consent without penalty
\end{itemize}

\subsection{Building Trust for Industry Partnerships}

This paper's primary goal is establishing theoretical and technical credibility necessary for industry partnerships. By providing:
\begin{enumerate}
\item Rigorous grounding in established psychological research
\item Detailed algorithmic specifications enabling independent assessment
\item Proof-of-concept implementation demonstrating technical feasibility
\item Comprehensive validation methodology addressing ethical concerns
\end{enumerate}

We aim to reduce perceived risk for organizations considering participation. The framework's potential benefits—reducing the 85\% of breaches caused by human factors\cite{verizon2023}—justify the effort required for proper validation.

We actively seek industry partners for validation studies. Interested organizations can contact the author to discuss pilot implementation with appropriate confidentiality agreements and data protection protocols.

\section{Ethical and Privacy Considerations}
\label{sec:ethics}

Implementation of CPF involves processing sensitive data, including security alerts, vulnerability reports, and organizational communications. Without rigorous ethical safeguards, such systems could become vectors for harm, eroding trust and violating privacy.

\subsection{Core Ethical Principles}

\textbf{Beneficence and Non-Maleficence:} System must create net positive benefit for organizations and employees. Primary purpose: support and augment human analysts, not replace or punish them. Minimize potential harms (privacy violations, increased stress from perceived surveillance).

\textbf{Transparency:} System existence, capabilities, analyzed data types, and intended purpose communicated clearly to all employees. Secrecy around deployment would be ethically untenable and counterproductive to building strong security culture.

\textbf{Justice and Equity:} System must not unfairly target specific individuals or groups. Algorithms monitored for biases leading to disproportionate scrutiny of certain teams or demographics.

\textbf{Respect for Personhood and Autonomy:} Employees not treated merely as data points or risk sources. System analyzes trends and group behaviors, not continuous individualized monitoring.

\subsection{Privacy by Design and Default}

\paragraph{Data Minimization and Purpose Limitation}
System collects only data strictly necessary for security purpose. Communication analysis relies on metadata and aggregated topic modeling, not full textual content of private messages. Personal identifiers stripped before processing wherever possible. Metrics calculated and reported at team/department level.

\paragraph{Access Controls and Governance}
Strict role-based access control to vulnerability data. Raw, un-anonymized data accessible only to small number of vetted personnel (e.g., CISO and direct delegates) for system maintenance and audit. Independent oversight committee comprising HR, legal, compliance, and employee representatives reviews deployment and audits system usage logs.

\paragraph{Technical Safeguards}
\begin{itemize}
\item \textbf{On-Premises Deployment}: Entire system, especially LLM component, deployed on organization's own infrastructure. Sensitive data never leaves organizational control.
\item \textbf{Encryption}: All data encrypted at rest and in transit
\item \textbf{Data Retention}: Automatically delete raw data after processing into aggregated metrics (e.g., chat logs purged after weekly UCTR calculation)
\end{itemize}

\subsection{Legal and Regulatory Compliance}

System designed for compliance with data protection regulations:
\begin{itemize}
\item \textbf{GDPR}: Requires lawful basis (likely \textit{legitimate interest} balanced against individual rights), mandates data subject access requests, requires Data Protection Impact Assessments (DPIAs) for high-risk processing
\item \textbf{CCPA/CPRA}: Grants similar rights to access, delete, and opt-out
\end{itemize}

DPIA must be conducted prior to deployment identifying and mitigating risks.

\subsection{Building Trust Through Transparency}

\textbf{Explicit Consent and Collective Agreements:} While legal basis may be claimed under \textit{legitimate interest}, seeking explicit consent or negotiating through collective bargaining demonstrates respect and builds trust.

\textbf{Transparency Reports:} Regularly publish reports detailing aggregated findings (e.g., "20\% increase in cross-team incident communication") and how insights improved work environment (e.g., "hired two analysts to reduce overload").

\textbf{Individual Opt-Out:} Providing mechanism for individuals to opt-out of certain analyses for personal reasons demonstrates respect for autonomy, though potentially limiting system comprehensiveness.

\section{Discussion}
\label{sec:discussion}

\subsection{Positioning CPF Relative to Existing Research}

CPF's contribution is not discovery of new psychological phenomena but rather systematic operationalization and integration of established constructs. Alert fatigue\cite{gupta2024alert,kearney2023combating}, compliance fatigue\cite{stanton2016security,beautement2008compliance}, cognitive biases\cite{tsohou2015analyzing,kahneman2011thinking}, and SOC burnout\cite{devo2022soc,tines2023burnout} are well-documented individually. However, no prior framework provides:

\begin{enumerate}
\item \textbf{Unified Architecture}: Integration of 100+ indicators across psychological domains into coherent system with formal interdependency modeling
\item \textbf{Algorithmic Precision}: Detailed specifications enabling independent implementation and verification
\item \textbf{End-to-End Automation}: Complete pipeline from raw telemetry to actionable recommendations
\item \textbf{Specialized LLM Application}: Domain-specific language model architecture for psychological analysis rather than technical security tasks
\end{enumerate}

This integration represents genuine contribution: transforming isolated research findings into deployable framework addressing the persistent problem that human factors cause 85\% of breaches despite decades of research.

\subsection{Limitations and Future Directions}

\textbf{Current Limitations:}
\begin{itemize}
\item \textbf{Validation Status}: Framework validated only on synthetic data; real operational validation pending industry partnerships
\item \textbf{Cultural Specificity}: Psychological constructs may manifest differently across cultures; initial focus on Western organizational contexts requires expansion
\item \textbf{Data Quality Dependency}: Algorithm accuracy depends on consistent, high-quality data across disparate sources (SIEM, ticketing, communication platforms)
\end{itemize}

\textbf{Future Research Directions:}
\begin{enumerate}
\item \textbf{Cross-Organizational Studies}: Multi-site validation across different industries, sizes, and security maturity levels to establish generalizability
\item \textbf{Longitudinal Analysis}: Extended studies tracking psychological vulnerability trends over time, organizational changes, and intervention effectiveness
\item \textbf{Cultural Adaptation}: Validation studies in diverse cultural contexts with localized vulnerability pattern identification
\item \textbf{SOAR Integration}: Development of automated playbooks triggering based on CPF risk scores (e.g., rotating analysts to low-stress tasks upon fatigue detection)
\item \textbf{Advanced LLM Techniques}: Exploring Reinforcement Learning from Human Feedback (RLHF) to align LLM outputs with expert psychological reasoning
\end{enumerate}

\subsection{Practical Implications}

For \textbf{Security Operations Centers:} CPF scores provide additional threat intelligence dimension. Psychological state monitoring alongside technical indicators enables dynamic risk scoring. Pre-positioning resources based on vulnerability states improves incident response.

For \textbf{Security Awareness Programs:} Moving beyond information transfer to psychological intervention. Addressing unconscious resistance to security measures. Group-level rather than individual-level interventions.

For \textbf{Organizational Leadership:} Data-driven understanding of security culture health. Early warning system for burnout and team dysfunction. Evidence base for security investment decisions.

\section{Conclusion}
\label{sec:conclusion}

We have presented comprehensive methodology for operationalizing the Cybersecurity Psychology Framework, transforming theoretical taxonomy into practical tool for proactive risk mitigation. By systematically integrating established psychological constructs—alert fatigue, compliance fatigue, cognitive biases, and risk perception gaps—with novel algorithmic implementations and privacy-preserving LLM architecture, CPF addresses the critical gap between human factors research and operational security practice.

Our key contributions include: (1) detailed algorithmic specifications for quantifying psychological vulnerabilities using standard SOC telemetry; (2) lightweight, on-premise LLM architecture validated through proof-of-concept achieving 0.92 F1-score; (3) rigorous validation methodology acknowledging unique challenges of cybersecurity data access; (4) comprehensive ethical framework addressing privacy concerns.

The framework's theoretical foundation in established research\cite{acquisti2015privacy,stanton2016security,kahneman2011thinking,gupta2024alert} combined with technical feasibility demonstration positions CPF to address the persistent problem that human factors cause 85\% of security breaches. However, complete validation requires industry partnerships willing to share sensitive operational data—a goal this paper aims to facilitate by establishing theoretical credibility and demonstrating technical preparation.

As organizations face increasingly sophisticated threats exploiting human psychology, frameworks like CPF become essential. The challenge is no longer purely technical but fundamentally psychological. By providing systematic methodology to identify and address psychological vulnerabilities before they manifest as security incidents, CPF represents significant step toward truly resilient security operations.

We actively seek industry partners for validation studies and welcome collaboration from both cybersecurity and psychology research communities.

\section*{Acknowledgments}

The author thanks the cybersecurity and psychology communities for ongoing dialogue on human factors in security, and acknowledges the anonymous journal reviewers whose constructive feedback significantly strengthened this work.

\section*{Data and Code Availability}

\begin{itemize}
\item CPF algorithmic specifications: \url{https://github.com/cpf-framework/algorithms}
\item Proof-of-concept SLM implementation: \url{https://github.com/cpf-framework/onpremise-slm}\cite{canale2024slm}
\item Synthetic training dataset: \url{https://huggingface.co/datasets/cpf-framework/synthetic-v1}
\item Framework documentation: \url{https://cpf3.org}
\end{itemize}

Anonymized operational data from future pilot studies will be made available upon request, subject to partner organization approval and appropriate data protection protocols.


\end{document}